\documentclass[10pt,conference]{IEEEtran}
\IEEEoverridecommandlockouts 
\usepackage{cite}
\newcommand{\citet}[1]{\cite{#1}}
\newcommand{\citep}[1]{\cite{#1}}

\usepackage{cite}
\usepackage{balance}
\usepackage{amsmath,amssymb,amsfonts}
\usepackage{algorithmic}

\usepackage{graphicx}

\usepackage{multirow}

\usepackage{textcomp}

\usepackage{xurl}

\usepackage{pdflscape}

\usepackage[many]{tcolorbox}

\usepackage{fontawesome}

\usepackage{hyperref}

\definecolor{main}{HTML}{CFCFCF}  
\definecolor{sub}{HTML}{CFCFCF}   

\tcbset{
    sharp corners,           
    colback = white,         
    before skip = 0.2cm,     
    after skip = 0.5cm       
}

\newtcolorbox{boxC}{
    colback = sub,  
    boxrule = 0pt   
}

\newcounter{keyTakeAwaysCounter} 
\def\BibTeX{{\rm B\kern-.05em{\sc i\kern-.025em b}\kern-.08em
    T\kern-.1667em\lower.7ex\hbox{E}\kern-.125emX}}

\AtBeginDocument{%
  \providecommand\BibTeX{{%
    Bib\TeX}}}

\usepackage{flushend}

\begin{document}

\title{The Invisible Hand of AI Libraries\\Shaping Open Source Projects and Communities}

\author{
\IEEEauthorblockN{
Matteo Esposito\textsuperscript{a}, 
Andrea Janes\textsuperscript{b}, 
Valentina Lenarduzzi\textsuperscript{a,c}, 
Davide Taibi\textsuperscript{a,c}
}
\IEEEauthorblockA{
\textsuperscript{a}University of Oulu, Finland, \textsuperscript{b}Free University of Bozen-Bolzano, Italy,\\\textsuperscript{c}University of Southern Denmark, Vejle, Denmark \\
matteo.esposito@oulu.fi, andrea.janes@unibz.it, valentina.lenarduzzi@oulu.fi, davide.taibi@oulu.fi
}
}

\maketitle

\begin{abstract}
In the early 1980s, Open Source Software emerged as a revolutionary concept amidst the dominance of proprietary software. What began as a revolutionary idea has now become the cornerstone of computer science. Amidst OSS projects, AI is increasing its presence and relevance. However, despite the growing popularity of AI, its adoption and impacts on OSS projects remain underexplored.

We aim to assess the adoption of AI libraries in Python and Java OSS projects and examine how they shape the development, i.e., technical ecosystem and community engagement. To this end, we will perform a large-scale analysis on 157.7k potential OSS repositories, employing repository metrics and software metrics to compare projects adopting AI libraries against those that do not. We expect to identify measurable differences in development activity, community engagement, and code complexity between OSS projects that adopt AI libraries and those that do not, offering evidence-based insights into how AI integration reshapes software development practices.
\end{abstract}


\section{Introduction}
In the early 1980s, Open Source Software (OSS) emerged as a revolutionary concept amidst the dominance of proprietary software. To keep the OSS momentum, Richard Stallman, in 1985, authored the GNU Manifesto \citep{stallman1985gnu}, marking a pivotal moment in the OSS movement. What began as a revolutionary idea \citep{doi:10.1080/08874417.2009.11646043} has now become the cornerstone of computer science \citep{1541831}, with OSS finding widespread adoption across both academic research and industry \citep{8737777}.
Artificial intelligence (AI) thrives within the realm of OSS. The widespread of libraries capable of managing sophisticated Machine Learning (ML) algorithms or designing intricate neural networks has accelerated advancements in both research and industry, paving the way to ``\textit{software 2.0}'' \citep{10.1145/3453478}. 

State-of-the-art commercial and OSS products expand their features daily by integrating AI libraries into OSS \citep{DBLP:journals/tse/YuBSM21}. AI libraries, comprising pre-built algorithms and frameworks, enable developers to implement AI functionalities in their projects seamlessly \citep{10274712}. AI libraries have democratized access to AI capabilities, enabling developers of varying skill levels to incorporate AI-driven features into their software projects without extensive machine learning or AI development expertise \citep{9695219}. Hence, it fosters inclusivity within the developer community, allowing individuals from diverse backgrounds to contribute to and benefit from AI-driven innovation \citep{10274712,9695219,DBLP:journals/tse/YuBSM21}. Developers leverage AI libraries as building blocks, accelerating the development cycle, improving governance, and fostering a culture of innovation within open-source projects \citep{9695219, esposito2025icseseip}.

In this context, our study will investigate the impact of adopting AI libraries in OSS. To our knowledge, no previous research has explored the impact of a broader spectrum of AI libraries on the OSS ecosystem, whether on a small or large scale. 

We aim to measure the extent of AI libraries adoption across Java and Python-based OSS and how it affects community engagement, development process, software complexity, and maintainability. Eventually, such a result would aid researchers and practitioners in improving their decision-making strategies when developing and managing AI-based OSS projects. 

We will perform a large-scale analysis of 157,7k potential repositories. Therefore, our study seeks to provide the following contributions:

\begin{enumerate}
    \item A comprehensive quantification of AI library adoption in OSS;
    \item Insights into development practices and community engagement in AI-based OSS;
    \item Code-Level analysis of structural, complexity, and maintainability affected by AI adoption.
\end{enumerate}

\section{Background \& Related Work}
One distinguishing feature of OSS is that its source code is openly accessible for anyone to use, modify, and distribute. In contrast to proprietary software developed and owned by individual companies, OSS fosters collaborative and community-driven development  Fortunato et al.~\citep{fortunato2021case}.  These characteristics make OSS a suitable candidate for diverse research in software engineering Gyimothy et al.~\citep{1542070}.
\subsection{OSS in Empirical Software Engineering}
Mockus et al.~\citet{10.1145/567793.567795}  compare commercial and open-source methodologies using Mozilla and Apache as examples. They find that while commercial projects rely on explicit coordination, OSS projects like Apache use decentralized communication. Mozilla faces challenges due to module interdependencies, while Apache's approach emphasizes modularity. Leveraging the open-source community for low-interdependence tasks can improve productivity. 

Similarly, Gyimothy et al.~\citet{1542070} conducted a study to assess the quality and reliability of OSS systems, mainly focusing on Mozilla, a widely used web and email suite. They calculated object-oriented metrics proposed by Chidamber et al.~\citet{chidamber1994metrics} to analyze fault-proneness in Mozilla's source code. 
Moreover, Yu et al.~\citet{DBLP:journals/tse/YuBSM21} conducted a large-scale empirical study on Java annotations, analyzing 1,094 OSS projects on GitHub. They investigated annotation usage, evolution, and impact, revealing 10 novel findings. Similarly, we are also interested in the context of OSS projects. Nonetheless, the author focused on Java annotations, while our study focuses on AI libraries and technologies. 

Arberdour et al.~\citet{4052554} highlights the significance of the OSS community's research on OSS quality, emphasizing the shift towards empirical data over subjective opinions. Moreover, they discuss the consensus in the literature on the key components of OSS quality. The authors review the state of the art, extracting lessons and exploring the divergences and potential inclusion of OSS quality approaches into closed-source software development methods.

Esposito et al.~\citet{esposito2023uncovering} investigated the hidden bugs, i.e., buggy untouched methods, that remained dormant for various releases of OSS projects.
 Similarly, Esposito et al.~\citet{esposito2023can} performed a large-scale analysis on OSS vulnerability severity according to SonarQube and the National Vulnerability  (NVD) Datasets to assess whether the NVD vulnerability severity was correlated with SQ severity levels. Finally, Esposito et al.~\citep{esposito2024extensive} performed a large-scale OSS-oriented test of static analysis security testing tools. Moreover, recent studies employed the ready availability of OSS development history to improve JIT defect prediction


\subsection{AI libraries in Software Engineering}
AI stimulated new research approaches to each software engineering task \citep{1701944}. 
Wan et al.~\citet{8812912}, investigated how incorporating machine learning into systems impacts software development practices. They conducted both qualitative interviews and a quantitative survey to identify differences between developing machine learning systems and traditional systems. The study revealed significant disparities in software engineering aspects such as requirements, design, testing, and process, as well as in work characteristics like skill variety, problem-solving, and task identity. 

Moran et al.~\citet{8374985} devised an automated approach called ReDraw to streamline the transformation of GUI mock-ups into code for user-facing software. ReDraw utilizes computer vision techniques and deep convolutional neural networks. A data-driven algorithm then generates a hierarchical GUI structure, enabling automatic assembly of prototype applications for the Android platform. 
Esposito et al.~\citet{esposito2024validate}, compiled a systematic dataset review on the usage of AI in the field of vulnerability prediction.

Furthermore, Liu et al.~\citet{10149441} conducted an empirical study on Stack Overflow to explore developers' challenges in finding suitable ML/DL libraries for AI tasks. Based on their findings, they proposed MLTaskKG, a task-oriented ML/DL library recommendation approach. MLTaskKG utilizes a knowledge graph to capture AI tasks, ML/DL models, implementations, repositories, and their relationships. By extracting knowledge from various sources, including ML/DL resource websites, papers, and frameworks, MLTaskKG recommends libraries matching developers' requirements. 

Xu et al.~\citet{10.1145/3487569}, address the challenge of turning conceptual ideas into code, especially when dealing with unfamiliar library APIs, by investigating the promise and challenges of using ML for code generation and retrieval within the PyCharm IDE. They developed a plugin for PyCharm that combines code generation and retrieval functionality and orchestrates virtual environments to collect user events. While qualitative surveys indicate positive developer experiences, quantitative results regarding increased productivity, code quality, and program correctness are inconclusive. The study identifies critical points for improving future machine learning-based code assistants and demonstrates developers' preferences between code generation and retrieval.

Finally, Wang et al.~\citet{9772253} conducted a 12-year Systematic Literature Review on 1,428 ML/DL-related SE papers published between 2009 and 2020. Their analysis highlighted the impacts, complexities, and challenges of applying ML/DL techniques to SE tasks. They categorized the rationales for ML/DL technique selection into five themes and examined issues related to reproducibility, replicability, and model choices in SE tasks.

\subsection{AI libraries in OSS}
Dilhara et al.~\citet{10.1145/3453478} presented a large-scale empirical study to examine the challenges developers face utilizing ML libraries in Software-2.0 development. Key findings include a significant increase in the adoption of ML libraries, various usage patterns, and unique challenges such as the binary incompatibility of trained ML models. Our study focuses on the broader AI spectrum of libraries, evaluating the impact of adopting them.

Tufano et al.~\citet{tufano2024unveiling} investigated how developers use Large Language Models, like OpenAI's ChatGPT, in OSS projects. Out of the 1,501 instances of potential ChatGPT usage in open-source projects, the authors narrowed them down to 467 actual ChatGPT instances. \citet{tufano2024unveiling} categorized these instances into 45 tasks developers automate using ChatGPT, hence helping developers understand how to use LLMs effectively and providing researchers with insight into tasks that can benefit and those that do not from automated solutions. Our study focuses on the usage of AI libraries, thus including LLMs, in OSS projects, intending to investigate the impact that the adoption of such libraries has on community engagement and the technical ecosystem. 
\section{Empirical Study Design}\label{sec:design}
Our empirical study design follows the guidelines defined by Wohlin et al. \citep{Wohlin2000}. In the upcoming sections, we detail the specific research questions and goals that drive our study and the procedures we used for data collection and analysis.

\subsection{Goal, Research Questions, and Hypothesis}
We formalized the \textbf{goal} of this study according to the Goal Question Metric (GQM) approach~\citep{Basili1994} as follows:

The goal of this study is to \textit{investigate} AI libraries adoption 
\textit{for the purpose of} evaluating their influence on software development, 
complexity, maintainability, and community engagement, 
\textit{from the point of view of} practitioners and researchers, 
\textit{in the context of} Open-Source Software projects.

Based on the goal mentioned above, we defined two main \textbf{Research Questions} (\textbf{$RQ_s$}), which serve as the primary focus of our investigation.

\begin{center}
\begin{boxC}
\textbf{RQ$_1$} How widespread is the adoption of AI-related libraries in open-source software projects?
\vspace{1em}
\begin{itemize}
    \item \textbf{RQ$_{1.1}$} Does AI adoption differ between Java and Python projects? 
\end{itemize}
\end{boxC}
\end{center}

This $RQ$ aims to measure the extent of AI library adoption in OSS projects. AI adoption captures \textbf{whether and how much} a project integrates AI features and capabilities through external components, i.e., libraries. According to~\citet{tarrega2020measuring,kula2018developers,pashchenko2018vulnerable}, we can analyze evidence of a project dependency on three levels:
\begin{itemize}
    \item \textbf{Dependency-level evidence:} importing or referencing AI/ML libraries (e.g., TensorFlow, PyTorch, Scikit-learn, Hugging Face Transformers).
    \item \textbf{Code-level evidence:} using classes, functions, or APIs belonging to these libraries.
    \item \textbf{Commit-level or release-level evidence:} identifying when AI libraries were introduced and whether they continue to be maintained or expanded.
\end{itemize}
In this first $RQ$, we aim to measure the dependency-level evidence, specifically whether the OSS project includes references to AI libraries in its list of dependencies. Furthermore, we are keen to focus on Java and Python because they represent two dominant yet contrasting ecosystems in software development~\cite{cass2025top-programming-languages}. Python is the de facto language for AI and machine learning, offering a rich ecosystem of libraries, including TensorFlow, PyTorch, and scikit-learn~\cite{li2025bridging,khandare2023analysis}. In contrast, Java is widely adopted in enterprise and large-scale software systems, where AI integration is emerging~\cite{Azul2025}. Studying both allows us to capture differences between experimentation-oriented (Python) and production-oriented (Java) environments. Other languages were excluded as they either lack mature AI ecosystems (e.g, Go)~\cite{JetBrainsGoML2023, JetBrainsGoTrends2024}, or they are more challenging, e.g., steeper learning curve, memory management complexity, that may affect adoption compared to higher-level languages  (e.g. C++)~\cite{GeeksforGeeksCPlusPlusML}, or are less representative of large, diverse open-source communities suitable for comparative analysis~\cite{li2025bridging}.

Therefore, we want to understand the extent of AI usage in OSS code by conjecturing the following  two \textbf{null}-hypotheses (\textbf{$H_{0}$}) and \textbf{alternative} hypotheses (\textbf{$H_1$}):
\begin{itemize}
    \item$\mathbf{H_{01}}$: \textit{The proportion of OSS projects relying on AI libraries is less than or equal ($\leq$) to 50\% of the sampled projects.}
    \item$\mathbf{H_{02}}$: \textit{The proportion of OSS projects relying on AI libraries is the same ($=$) between Java and Python.}

    \item$\mathbf{H_{11}}$: \textit{The proportion of OSS projects relying on AI libraries is greater than ($>$) to 50\% of the sampled projects.}
    \item $\mathbf{H_{12}}$: \textit{The proportion of OSS projects relying on AI libraries is different ($\neq$) between Java and Python.}
\end{itemize}

Moreover, investigating how they shape the evolution and functionality of OSS projects offers novel perspectives on the challenges and impact of AI technology in OSS. Hence, we ask:

\begin{boxC}{\textbf{RQ$_2$}:}
 How does adopting AI libraries affect OSS projects' structural, collaborative, and technological characteristics?

 \begin{itemize}
    \item \textbf{RQ$_{2.1}$}: How does the use of AI libraries influence development activity, collaboration, and workflow practices in repositories?
    
    \item \textbf{RQ$_{2.2}$}: How does AI library usage affect the dynamics of community engagement and issue/pull request management in open-source repositories?

    \item \textbf{RQ$_{2.3}$}: Do repositories using AI libraries exhibit distinct patterns in release strategies, technology stacks, and automation practices?

 \end{itemize}
\end{boxC} 
In recent years, OSS projects have increasingly embraced AI libraries like TensorFlow, PyTorch, and Hugging Face to build smarter, data-driven applications covering a wide range of possible application fields spanning from innovative products to code generation and co-architecting complex software solutions \cite{chiu2023impact,makridakis2017forthcoming,dwivedi2021artificial,esposito_generative_2025}. Albeit such libraries bring powerful capabilities, they may also change how teams work, communities engage, and technology choices are made \cite{akbar_6gsoft_2024}.
While AI libraries are becoming a staple in modern development, we still know surprisingly little about how they shape the projects that adopt them. Do they influence how developers collaborate? Do they affect how often code is released or how issues and pull requests are handled? Understanding whether adopting AI libraries may shape some project aspects is crucial, especially as more OSS communities move toward AI-enabled development \cite{lagrandeur2023consequences}. Therefore, we aim to explore how adopting AI libraries impacts OSS projects across three key dimensions: how teams collaborate, how the community engages, and how technologies and workflows evolve. Following previous studies, we collected \textbf{repository metrics} (RMs) that provide valuable insights into the activity, health, and community engagement of a software project hosted on platforms like GitHub to measure such key aspects. Due to space constraints, we present in \textbf{Table I (online appendix)} the collectable RMs, their descriptions, and their classifications according to \cite{d2010extensive,bavota2015four,bugayenko2024cam,github_docs}. Based on our RQ goal and the RM metrics, we conjecture the following seven \textbf{null}-hypotheses (\textbf{$H_{0}$}) and \textbf{alternative} hypotheses (\textbf{$H_1$}):

\begin{itemize}

    \item $H_{03}$: \textit{AI library adoption does not affect development activity, collaboration, and workflow-related metrics.}
    \item $H_{04}$: \textit{AI library adoption does not affect community engagement and issue/pull-related metrics.}
    \item $H_{05}$: \textit{AI library adoption does not affect release-management, technology, and automation practices.}
    \item $H_{13}$: \textit{AI library adoption affects development activity, collaboration, and workflow-related metrics.}
    \item $H_{14}$: \textit{AI library adoption affects community engagement and issue/pull-related metrics.}
    \item $H_{15}$: \textit{AI library adoption affects release-management, technology, and automation practices.}
\end{itemize}

Moreover, there’s growing interest in understanding how their adoption affects the code itself, i.e., not just what the software accomplishes but how it’s structured, how complex it becomes, and how challenging it is to maintain. Hence, we ask:

\begin{boxC}{\textbf{RQ$_3$}:}
How does adopting AI libraries influence the internal structure, complexity, and maintainability of OSS projects at the code level?

 \begin{itemize}
    \item \textbf{RQ$_{3.1}$}: How does the use of AI libraries affect class design, inheritance, and coupling in open-source software systems?

    \item \textbf{RQ$_{3.2}$}: To what extent does AI library usage impact complexity and cohesion at the function and method level?

    \item \textbf{RQ$_{3.3}$}: How does adopting AI libraries influence code size, and overall maintainability?

 \end{itemize}
\end{boxC} 
AI libraries often bring new architectural choices, advanced abstractions, and dense logic, which may shape development practices in subtle but impactful ways \cite{esposito2024leveraging}.
While much attention has been paid to the capabilities these libraries enable, we still know little about their side effects on internal software quality. Do they make code more complex? Less cohesive? Do they change the way developers design and organize classes? 

With open-source repositories playing a central role in shaping modern AI-enabled applications, these questions are more relevant than ever.
With this study, we aim to explore whether, and how, the inclusion of AI libraries influences structural, complexity-related, and maintainability aspects of source code in open-source projects. To do this, we group our analysis around three core dimensions: (i) object-oriented design structure, (ii) control-flow complexity and cohesion, and (iii) maintainability and readability of the codebase. The goal is to identify statistically significant patterns that distinguish AI-enhanced projects from traditional ones at a deeper, code-centric level. Hence, to measure changes in such dimensions, we opted to employ \textbf{software product and quality metrics} (SMs) to provide quantitative insights into various aspects of software design, implementation, and maintainability~\cite{esposito2023uncovering}, obtainable via Understand~\cite{scitools_understand_2025} (Table II in the Online Appendix). Based on our RQ goal and the SMs, we conjecture the following three \textbf{null}-hypotheses (\textbf{$H_{0}$}) and \textbf{alternative} hypotheses (\textbf{$H_1$}):
\begin{itemize}
    \item \textbf{$H_{06}$}: AI library adoption does not affect class-level coupling, inheritance depth, or object-oriented design structure.  
    \item \textbf{$H_{07}$}: AI library adoption does not affect cyclomatic complexity or cohesion metrics.  
    \item \textbf{$H_{08}$}: AI library adoption does not affect code volume or comment density.  
    \item \textbf{$H_{16}$}: AI library adoption affects class-level coupling, inheritance depth, or object-oriented design structure.  
    \item \textbf{$H_{17}$}: AI library adoption affects cyclomatic complexity or cohesion metrics.  
    \item \textbf{$H_{18}$}: AI library adoption affects code volume or comment density. 

\end{itemize}

\subsection{Study Context}

The project selection process focused on GitHub repositories sourced from various owners. Two primary considerations have driven our choices. 

Firstly, GitHub stands out as the preeminent online source-code repository, providing a vast data repository for analysis. Therefore, it enables us to explore various projects and extract relevant data for our study. Secondly, we deliberately avoided focusing on specific owners to mitigate biases in library adoption practices across projects. Focusing solely on a single owner could skew results, as it might lead to standardized contribution practices and potentially uniform design choices across projects \citep{9273270}. Major repository owners such as the Apache Software Foundation often enforce contribution standards, resulting in a proliferation of similar design decisions and potential issues \citep{9273270,4375249}.

\section{Execution Plan}

\subsection{Study Setup and Data Collection}

\begin{figure*}[t]
    \centering
    \includegraphics[width=1\linewidth]{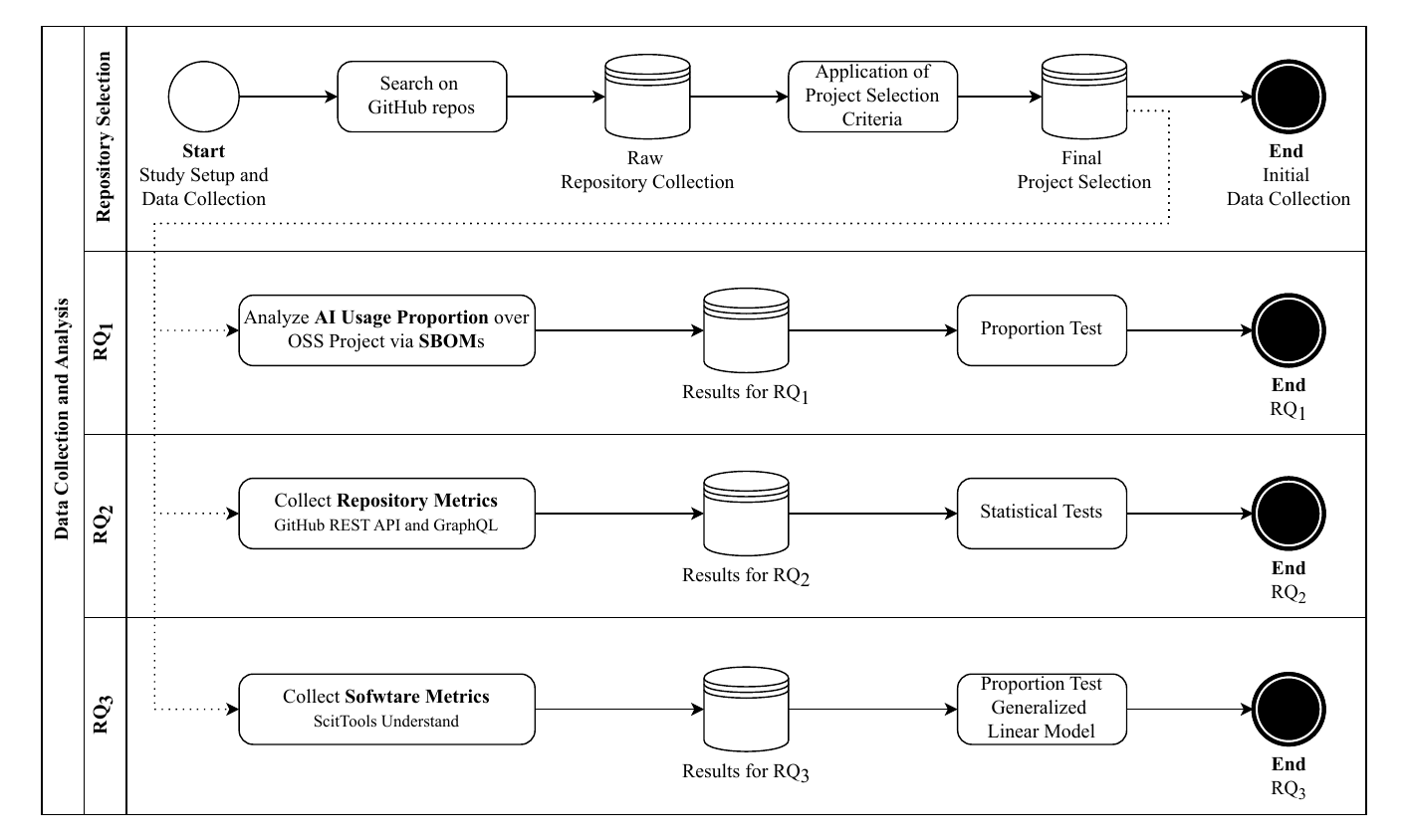}
    \caption{Data Collection and Analysis Workflow}
    \label{fig:DAWF}
\end{figure*}

This section describes our data collection methodology. Figure \ref{fig:DAWF} depicts our data collection workflow that utilizes the BPMN 2.0 notation \cite{object_management_group_business_2011}. According to Figure \ref{fig:DAWF}, our approach starts by programmatically accessing GitHub repositories to gather the raw data required for our study. We will employ GitHub Search and Repository APIs, performing two independent searches requesting repositories with Java or Python as their primary programming language. 

\subsubsection{GitHub Project Selection} The GitHub repository API offers a seamless way to interact with repositories without downloading them. 
Similar to \citet{dabic2021sampling}, we requested all repositories created in a specific time to overcome the API limitation. We define a dynamic time window. We will analyze the repositories created between the 1st of January 2000, and the data collection date. We default to using a time window of five years. In the case of precisely 1000 results, we reduced the window to one year, six months, one month, and one week~\cite{dabic2021sampling}. 
\cite{10.1145/3453478, DBLP:journals/tse/YuBSM21} inspired our selection process. 
Table \ref{tab:repository_criteria} summarizes the inclusion criteria for repositories. According to Table \ref{tab:repository_criteria}, we will retrieve all repositories whose star count (SC) will be greater than 50 when performing the full study~\cite{10.1145/3453478, DBLP:journals/tse/YuBSM21}. We employ SC as a proxy metric for repository popularity according to \cite{BORGES2018112}.  We will include only non-forked and unarchived projects and avoid inactive ones by discarding those that do not have a single activity event within six months before our data collection date \cite{10.1145/3453478}. Moreover, we included only projects that allowed access to the Software Bill of Materials\footnote{\url{https://docs.github.com/en/code-security/supply-chain-security/understanding-your-software-supply-chain/about-the-dependency-graph}} (SBOM) to obtain the dependency list. Finally, according to \cite{BORGES2018112}, we exclude repositories with fewer than three contributors.
\begin{table}
    \centering
    \caption{Repository Inclusion Criteria}
    
    \footnotesize
    \resizebox{\linewidth}{!}{%
    \begin{tabular}{p{2.5cm}p{5cm}}
\hline  
    \textbf{Criterion} & \textbf{Condition} \\
\hline  
    Archived or Fork & Repository is neither archived nor a fork \\
    Last Activity & The last activity on the repository is less than six months old \\
    Contributors & Repository has at least three contributors \\
    Star Count & Repository has 50 or more stars \\
    SBOM Availability & The GitHub repository allows access to the SBOM \\
\hline  
    \end{tabular}
    }
    \label{tab:repository_criteria}
\end{table}
Our final repository datasets, based on Java and Python repositories, will be as diverse as possible across application domains, size, development frameworks, contribution governance, and testing methodologies. In general, our data set diversity is paramount for addressing generalizability issues. 
Moreover, we are not excluding younger projects even though they may exhibit more instability regarding community engagement, development, and technology adoption. The topic has given rise to numerous recent yet highly active projects frequently adopted in production.

We performed a preliminary search on GitHub to evaluate the size of the project sample, only applying the criteria in Table \ref{tab:repository_criteria} to obtain the repositories of \textbf{40.7k Java} and \textbf{117.7k} Python repositories. Therefore, our evaluation presents a total of \textbf{157.7k} repositories as potential candidates for mining.

\subsubsection{AI Libraries Adoption ($RQ_1$)}
\label{subsubsec:sm}
Leveraging the final project selection, we collect the dependencies list to measure the adoption of AI libraries in OSS projects ($RQ_1$). More specifically, we will leverage the SBOMs, a structured inventory that provides comprehensive details about a software project's components, to analyze the adoption of AI libraries. It includes component names, versions, dependencies, licensing details, and sometimes security vulnerabilities \cite{10.1145/3453478}.  Parsing the SBOM files allows us to gather the list of all dependencies of each repository, enabling our analysis.

\subsubsection{Selection of AI Libraries}
   Due to a lack of standardization and automation in labeling third-party libraries in Python and Java, we create the selected AI library list by mining the two main library repositories: \textit{PyPI} for Python and \textit{Maven Central} for Java. 
    Regarding \textbf{PyPI}, we query both via the \textit{Trove Classifiers} labeled as ``Scientific/Engineering: Artificial Intelligence'' and extend the search using domain-specific keywords such as ``machine learning,'' ``deep learning,'' ``neural,'' ``transformer,''``large language model'',  and  ``generative ai''. 
    For \textbf{Maven}, we issue the same keyword-based queries over artifact names, group identifiers, and descriptions to capture Java-based AI libraries.

For each candidate library, we collect popularity metrics as a proxy of widespread adoption~\cite{souza2008empirical,dilhara2021understanding}: for PyPI, we collect download counts; for Maven Central, when available, \textit{download} and \textit{reverse-dependency counts}. In the case that Maven libraries do not have such information, we collect from the library's GitHub repository the \textit{stars count} metric. We compute the interquartile range (IQR) for the primary popularity metrics (stars for GitHub, downloads for registries) and retain libraries at or above $Q_3$ therefore removing long-tail entries with limited adoption, yielding a focused set of potentially widely used AI libraries for analysis.

Finally, we \textbf{manually validate} the shortlisted libraries to confirm their genuine AI orientation. 
Two authors will independently examine documentation, README files, and dependency structures to verify that each library actually provides AI features, thereby excluding ``AI-assisting'' libraries such as NumPy or Pandas that may be associated with AI workflows but do not provide any actual AI features. We measure the agreement via Cohen's $\kappa$~\cite{cohen_coefficient_1960} and solve disagreements involving the third author. Finally, for each AI library, we collect its package identifiers and match against the SBOM for each project.

\subsubsection{Community Engagement \& Development Process ($RQ_2$)}
Leveraging the final project selection, we will collect via  GitHub REST and GraphQL APIs, all the RMs (see Online Appendix) to measure the development activities ($RQ_2$). We will gather metrics on development activity, collaboration patterns, workflow automation, community engagement, issue and pull request management, technology stacks, and release strategies.

\subsubsection{Complexity and Maintainability ($RQ_3$)}

To investigate how adopting AI libraries influences the internal structure and maintainability of open-source projects at the code level ($RQ_3$), we will collect SMs using SciTools Understand, a widely-used static code analysis tool \cite{scitools_understand_2025}.

Understand has been frequently employed in software engineering literature due to its ability to reliably analyze large codebases across multiple programming languages, including Java and Python. For our analysis, we will locally clone all the repositories in the final project selection and extract class-, function-, and method-level metrics covering key dimensions such as inheritance, coupling, complexity, cohesion, and code size,i.e., all available Software Metrics, in terms of product, process, complexity and quality, measaured by Understand~\footnote{\url{https://scitools.com/metrics}}.

\subsection{Data Analysis}
\label{sec:DataAnalysis}

Since this is a registered report, we did not perform the study, so we cannot know ``a priori'' whether the data will be distributed normally. Therefore, in the rest of this section, we provide the data analysis protocol for both cases. 
\subsubsection{AI Libraries Adoption ($RQ_1$)}
To answer $RQ_1$, we will determine the proportion of projects using AI libraries based on the presence of popular AI libraries in the dependency list of the OSS projects in the SBOMs.  Since our quantitative analysis includes project selection, we cannot conclude by relying solely on a single observed proportion or descriptive statistic because it does not account for variability in the data or provide a measure of uncertainty.  

To determine whether the observed proportion exceeds a specified threshold ($H_{11}$) or whether the differences between Java and Python projects are statistically significant ($H_{12}$), we need to statistically test the hypotheses, since any observed proportion or difference could be misinterpreted as meaningful, even if it results from random sampling noise or biases in the data set.  The choice of statistical test depends on the normality of the sampling distribution of proportions. For proportions, normality is assessed by ensuring that the sample satisfies \(n \cdot p_0 \geq 10\) and \(n \cdot (1 - p_0) \geq 10\), where \(n\) is the sample size and \(p_0\) is the hypothesized proportion. If these conditions are met, we apply z-tests. Regarding $H_{11}$, a one-sample z-test evaluates whether the proportion of AI-adopting projects exceeds a threshold (e.g. 50\%), using the test statistic \(z = \frac{p_{\text{obs}} - p_0}{\sqrt{\frac{p_0(1-p_0)}{n}}}\), where \(p_{\text{obs}}\) is the observed proportion. Regarding $H_{12}$, a two-sample z-test compares AI adoption rates between Java and Python projects, with the test statistic given by \(z = \frac{(p_1 - p_2)}{\sqrt{p(1-p)\left(\frac{1}{n_1} + \frac{1}{n_2}\right)}}\), where \(p_1\) and \(p_2\) are the observed proportions for Java and Python, and \(p\) is the pooled proportion. 

When normal conditions are unsatisfactory, the exact methods provide more reliable results. Regarding $H_{11}$, we use the binomial test, which calculates the exact probability of observing at least \(x\) successes in the trials \(n\) under the null hypothesis \(p_0\). This is given by \(P(X \geq x) = \sum_{k=x}^{n} \binom{n}{k} p_0^k (1 - p_0)^{n-k}\), where \(\binom{n}{k}\) is the binomial coefficient. In contrast, for $H_{12}$, we use Fisher’s Exact Test to compare the distributions of AI adoption between Java and Python projects. This test evaluates the exact probability of the observed contingency table or a more extreme one, using the formula \(P = \frac{\binom{n_1}{a} \binom{n_2}{c}}{\binom{n_1 + n_2}{a+c}}\), where \(a\) and \(c\) are the counts of AI-adopting projects for Java and Python, respectively, and \(n_1\) and \(n_2\) are the total projects for each group. 

\subsubsection{Community Engagement \& Development Process ($RQ_2$)}
To answer $RQ_2$, we must test the distribution of the collected RMs. All RMs are continuos numerical variables, hence Shapiro-Wilk or Anderson Darling tests are suitable for testing for normality.  Therefore we conjecture one \textbf{hypothesis} ($H_\mathcal{N}$) as follows:

\begin{itemize}
    \item $H_{1\mathcal{NRM}}$: \textit{Repository metrics do not follow a normal distribution.}
\end{itemize}

\noindent Hence, we defined the \textbf{null hypothesis} ($H_0$) as follows:
\begin{itemize}
    \item $H_{0\mathcal{N}RM}$: \textit{Repository metrics follow a normal distribution.}
\end{itemize} 

We tested $H_\mathcal{NRM}$ with the Anderson-Darling (AD) test \citep{andersondarling1952}. The AD test assesses whether data samples derive from a specific probability distribution, such as the normal distribution. AD measures the difference between the sample data and the expected values from the tested distribution. More specifically, it evaluates differences in the cumulative distribution function (CDF)  between the observed data and the hypothesized distribution \citep{andersondarling1952}.

The AD  and the Shapiro-Wilk (SW)  \citep{shaphiro1965analysis} are both statistical tests used to assess the normality of data.  The SW test focuses on the correlation between the observed data and the expected values under a normal distribution, emphasizing the smallest and largest values in the dataset. Therefore, according to \citet{mishra2019descriptive}, SW is more appropriate when targeting small datasets ($\leq50$ samples). 
On the other hand, the AD test considers a broader range of values, including those in the middle of the distribution, providing a more sensitive evaluation of normality, especially for larger sample sizes than our own. 
Therefore, we preferred  AD over SW \citep{Stephens1974EDF}. 
AD is also considered one of the most powerful statistical tools for detecting most departures from normality \citep{Stephens1974EDF, robredo_manero_analyzing_2024,esposito_call_2025}.

In case of \textbf{acceptance of the null hypotheses ($H_\mathcal{NRM}$)}, similarly to the normality testing sample concerns, we will test $H_{1(3\rightarrow5)}$  with the Z-statistic \citep{sprinthall2011basic}. We can use the Z-statistic to assess the significance of observed differences between sample data and population parameters. Calculated by standardizing sample statistics based on known population parameters, the Z-statistic quantifies the extent to which observed results deviate from what would be expected under the null hypothesis \citep{sprinthall2011basic,casella2002statistical,montgomery2020applied}. 

Interpreting the Z-statistic involves comparing its value with critical values of the standard normal distribution, typically determined by the chosen significance level \citep{casella2002statistical}. If the calculated Z statistic falls within the critical region, it suggests that the observed results are extreme enough to reject the null hypothesis in favor of the alternative \citep{montgomery2020applied}. 

In case of \textbf{rejection of the null hypothesis ($H_\mathcal{N}$)}, we will test $H_{1(3\rightarrow5)}$  using the Wilcoxon signed rank test (WT) \citep{pub.1102728208}, which is a nonparametric statistical test that compares two related samples or paired data. WT uses the absolute difference between the two observations to classify and then compare the sum of the positive and negative differences. The test statistic is the lowest of both. We selected WT to test $H_{1(3\rightarrow8)}$  because RMs' observations would not be normally distributed; hence, we used it instead of the paired t-test, which assumes a normal data distribution. 

We set our alpha to 0.01. We decided to reduce the alpha value from the standard value of 0.05 due to the many statistical tests we will perform and therefore to reach a good balance between Type I and II error \citep{abac.12172}.

\subsubsection{Complexity and Maintainability ($RQ_3$)}
To answer $RQ_3$, we must assess the distribution of the collected Software Metrics (SMs). All SMs are continuous numerical variables, making them suitable for normality testing using AD. We define the following \textbf{hypothesis} ($H_\mathcal{N}$):

\begin{itemize}
\item $H_{1\mathcal{N}SM}$: \textit{Software metrics do not follow a normal distribution.}
\end{itemize}

\noindent Accordingly, the \textbf{null hypothesis} ($H_0$) is defined as:

\begin{itemize}
\item $H_{0\mathcal{N}SM}$: \textit{Software metrics follow a normal distribution.}
\end{itemize}

Similar to $RQ_2$, in the case of \textbf{acceptance of the null hypothesis} ($H_{0\mathcal{N}SM}$), we will test the hypothesis $H_{1(6\rightarrow8)}$ using the Z-statistic \citep{sprinthall2011basic}. 

However, in the case of \textbf{rejection of the null hypothesis} ($H_{0\mathcal{N}SM}$), we will instead adopt the Wilcoxon signed-rank test (WT) \citep{pub.1102728208} to evaluate $H_{1(6\rightarrow8)}$.

\subsection{Replicability and Verifiability}
\label{sec:Replicability}
Our work will adhere to the \textbf{Open Science} principles\cite{open_science_principles}. We will prepare the replication package, containing raw data and scripts, and share it on Zenodo, according to \textbf{FAIR} principles \cite{force11_fair}, i.e., Findable, Accessible, Interoperable, and Reusable to foster applicability and verifiability and bolster the feasibility of future works.

\section{Threats to Validity}\label{sec:threats}
In this section, we discuss the threats to the validity of our study. We categorized the threats into Construct, Internal, External, and Conclusion, following the guidelines defined by Wohlin et al.~\citep{DBLP:books/daglib/0029933}.

\textbf{Construct Validity}. Our specific design choices, including our measurement process and data filtering, may impact our results. To address this threat, we based our choice on past studies \citep{dabic2021sampling,tufano2024unveiling,10.1145/3453478,DBLP:journals/tse/YuBSM21} and used well-established guidelines in designing our methodology \citep{DBLP:journals/ese/RunesonH09,Basili1994}.

\textbf{Internal Validity}.  Our study relies on a large-scale analysis of 6,323 repositories, which can potentially be biased from the project selection. We address this threat by designing our inclusion criteria based on past studies  \citet{10.1145/3453478, DBLP:journals/tse/YuBSM21, dabic2021sampling}. 

\textbf{External Validity}. 
Mining versioning systems, particularly GitHub, also threaten external validity. More specifically, GitHub's user base predominantly comprises developers and contributors to open-source projects, potentially skewing findings towards this specific demographic. In our context, this is not a real threat because it is the study's focus. 
Moreover, the dynamic nature of GitHub, with frequent updates, forks, and merges, poses challenges in ensuring the stability and consistency of data over time. We addressed this issue by providing the raw data in our replication package. Furthermore, the accessibility of GitHub data is subject to various permissions and restrictions set by project owners, potentially hindering reproducibility and transparency in research. We addressed this issue, limiting our data collection to public repositories. 
Nonetheless, the vast diversity of projects analyzed in our study presents a significant diversity in terms of programming languages, project sizes, and development methodologies, thus aiding the generalizability of results. 

\textbf{Conclusion Validity}. We do not discuss the validity of the conclusion since this is a registered report. 
\section{Conclusions}\label{sec:conclusions}
Our registered report sets the foundation for a large-scale empirical investigation into the adoption of AI libraries and how they affect the OSS development and ecosystem. Upon completion, we expect our study to yield actionable insights into (i) the prevalence of AI adoption, (ii) its impact on community engagement and development practices, and (iii) its influence on code complexity and maintainability. The results of this study are expected to contribute significantly to both research and practice. 

To researchers, our findings will provide empirical grounds for theorizing on how AI library adoption transforms software development practices, community behaviors, and internal code quality, opening up new venues in socio-technical dynamics, AI-facilitated development, and software maintainability. 
Concerning developers, we will provide insights and recommendations from the OSS practices of incorporating AI libraries into software projects, allowing teams to make informed decisions on tool adoption, development practices, and maintenance in the long run.

 \section*{Acknowledgments}
This work has been funded by the Research Council of Finland (grants n. 359861 and 349488 - MuFAno) and Business Finland (grant 6GSoft\cite{akbar_6gsoft_2024}).
\section*{Online Appendix}
Due to space constraints, we provide the complete tables of all repositories, Understand metrics, and the methodology of the confounding-factor controls for RQ$_2$ and RQ$_3$ in the online appendix of this registered report\footnote{\url{https://doi.org/10.5281/zenodo.17557533}}.

\balance
\bibliographystyle{IEEEtran}
\bibliography{main}
\end{document}